\documentclass{IEEEtran}
\AtBeginDocument{%
  }

\usepackage{graphicx} 
\usepackage{tabularx}
\usepackage{booktabs}
\usepackage{adjustbox}
\usepackage{framed}
\usepackage{makecell}
\usepackage{url}
\usepackage{color,soul}
\usepackage{booktabs}
\usepackage{multirow} 
\usepackage{algorithmic}
\usepackage{textcomp}
\usepackage{xcolor}
\usepackage{todonotes}
\usepackage{listings}
\usepackage{wrapfig}
\usepackage{tabularx}
\usepackage{caption}
\usepackage{subcaption}

\usepackage[framemethod=TikZ]{mdframed}
\definecolor{boxcolor}{HTML}{6e44ff}
\newenvironment{coloredframe}[2][]{
    \mdfsetup{
    skipabove=3pt, skipbelow=3pt,
    innerlinewidth=1.5pt, 
    innerlinecolor=white, 
    linewidth=0pt,
    backgroundcolor=#2!10,
    innerleftmargin=10pt, 
    innerrightmargin=15pt,      
    innertopmargin=10pt,        
    innerbottommargin=10pt,     %
    roundcorner=5pt           
    }
    \begin{mdframed}}
    {\end{mdframed}}

\begin{document}

\title{From Human-to-Human to Human-to-Bot Conversations in Software Engineering}

\author{\IEEEauthorblockN{Ranim Khojah, Francisco Gomes de Oliveira Neto, Philipp Leitner}
\IEEEauthorblockA{ \\
\textit{Chalmers $\vert$ University of Gothenburg}\\
Gothenburg, Sweden \\
khojah@chalmers.se, francisco.gomes@cse.gu.se, philipp.leitner@chalmers.se}
}

\maketitle
\begin{abstract}
    Software developers use natural language to interact not only with other humans, but increasingly also with chatbots. These interactions have different properties and flow differently based on what goal the developer wants to achieve and who they interact with.
    In this paper, we aim to understand the dynamics of conversations that occur during modern software development after the integration of AI and chatbots, enabling a deeper recognition of the advantages and disadvantages of including chatbot interactions in addition to human conversations in collaborative work.
    We compile existing conversation attributes with humans and NLU-based chatbots and adapt them to the context of software development. Then, we extend the comparison to include LLM-powered chatbots based on an observational study. We present similarities and differences between human-to-human and human-to-bot conversations, also distinguishing between NLU- and LLM-based chatbots. 
    Furthermore, we discuss how understanding the differences among the conversation styles guides the developer on how to shape their expectations from a conversation and consequently support the communication within a software team. We conclude that the recent conversation styles that we observe with LLM-chatbots can not replace conversations with humans due to certain attributes regarding social aspects despite their ability to support productivity and decrease the developers' mental load.

\end{abstract}

\begin{IEEEkeywords}
    Conversational agents, Software development, Dialogue characteristics
\end{IEEEkeywords}

\section{Introduction}

In software development, discussions and conversations within a team play an important role in communicating progress and resolving issues.
These conversations are not only between humans --- in modern software projects, developers also interact with chatbots through conversational interfaces, and increasingly also with powerful generative AI models, such as Large Language Models (LLMs). LLMs in particular can be seen as a hype topic, with multiple recent studies investigating their usage in software development~\cite{ross2023programmer,ipek2023applications}.

Intuitively, despite using the same interface (natural language), conversations with chatbots and LLMs follow different rules and have different purposes and constraints than conversations with fellow humans. So far, these differences are ill-understood, with developers either anthropomorphizing interactions with smart programming tools, or underusing their capabilities.

In this article, we provide a structured comparison of conversations between: (i) humans, (ii) humans and chatbots based on natural language understanding (NLU), and (iii) humans and LLM-based chatbots. Following a comparison taxonomy by Clark et al.~\cite{clark2019makes}, we discuss differences in purpose, understanding, trustworthiness, listening, and use of humour. A particular focus of our work is LLM-based chatbots, and their place in modern software development processes.

Our discussion aims to assist software engineers in calibrating their expectations when engaging with different conversational partners, be they other humans or various types of chatbots. By understanding the dynamics of these interactions practitioners can better recognize the potential gains and losses of collaborative work when substituting human dialogue with chatbot interactions, particularly in the realms of knowledge sharing.

\section{A Research View on Bots in Software Development}
\label{sb:related-work}

Bots in software development (DevBots) have become common tools that play a big role in improving developers' productivity and communication throughout the development process.
With the constant emergence of new DevBots each day, research was first focused on understanding the role of bots. Erlenhov et al. \cite{erlenhov2019current} focus on the DevBots roles and introduce a taxonomy to classify DevBots according to their functionalities and how they contribute to enhancing software development processes. While Wessel et al. \cite{wessel2018power} further explore the use cases of DevBots in open-source software and found that they are mostly used for automation purposes. 
Among the various types of DevBots, chatbots have drawn more attention due to their ability to communicate with developers through natural language, expanding the spectrum of tasks that DevBots can assist with.

Earlier chatbots were powered by natural language understanding (NLU) to automate repetitive and simple tasks. For example, MSRBot \cite{abdellatif2020msrbot} that answers questions related to a software repository, and Stack Overflow Bot\footnote{\url{ https://aka.ms/stackoverflow-bot-sample}} has been used to retrieve relevant information from Stack Overflow. 
NLU-based chatbots operate by predicting the user's intention and then mapping it to a corresponding response in the chatbot's database. Therefore, the possible use cases of such chatbots are limited to the database and training data. 

More recently, Large Language Models (LLMs) introduced a generative aspect to chatbots, making it possible for chatbots to generate responses to queries it was not trained on. This revealed more possible use cases that can be more complex due to their need to understand, synthesize, and create artifacts. For example, GitHub Copilot\footnote{\url{http://copilot.github.com}} generates code given the user's specifications and the context of the project.
The use cases were not confined to code generation but also testing, requirement analysis, and other activities.

To evaluate the effectiveness of chatbots and how they can support software engineers, current research focuses on the quality \cite{clark2019makes}, usability \cite{lee2021don}, and helpfulness \cite{yang2019understanding} of the chatbot's outcome. Nevertheless, we lack a holistic understanding of the experience and the flow of the conversation between software engineers and chatbots.

\section{An Observational Study}
\label{sb:fse}

The reflections and insights of this paper are based on a dataset and findings from our previous work~\cite{khojah2024beyond}. The study aimed to understand how software engineers use ChatGPT in their workplace in real-world settings. Particularly, the previous paper analyzed the user experience of ChatGPT, focusing on practitioners' goals when interacting with ChatGPT and the helpfulness of dialogue outcomes. In contrast, this study distinguishes the different ways conversations occur between humans, as well as between humans and various types of bots, using data from previous work to particularly highlight contrasts in human and LLM-based chatbot interactions.

To collect the data for this study, we reached out to practitioners in different software organizations that allow the use of ChatGPT in their workplaces. In total, 24 software engineers from 10 software organizations of different sizes and domains registered to participate in our study (See Table \ref{tab:org-info}). The data is available in Zenodo~\cite{khojah2023replication} and includes information about the purpose of each prompt from participants and how the prompts support various Software Engineering tasks such as coding, testing, and design. However, we cannot share the chat files or the open-ended survey responses, as they might compromise participant anonymity or contain company-specific information.

\begin{table}[!ht]
    \centering
    \caption{Information about how the participants are distributed across different organizations of different sizes and domains. We refer to each organization with an ID, and the sizes used are Startups, Small and Medium enterprises (SME), and Large enterprises.}
    \label{tab:org-info}
    \begin{tabularx}{\linewidth}{lXXr}
    \toprule
    \textbf{Org. ID} & \textbf{Org. Size} & \textbf{Domain} & \textbf{\# Participants} \\
    \midrule
    A & SME     & Testing     & 3 \\
    B & SME     & E-learning  & 3 \\
    C & Startup & Medical     & 4 \\
    D & Startup & Gaming      & 1 \\
    E & Startup & Gaming      & 1 \\
    F & Large   & E-commerce  & 1 \\
    G & Large   & Automotive  & 7 \\
    H & Large   & Consultancy & 1 \\
    I &  SME    & Consultancy & 2 \\
    J &  Large  & Automotive  & 1 \\
    \bottomrule
\end{tabularx}
\end{table}

The participants actively used ChatGPT for a week during their normal work, that is, to perform tasks that are relevant to their role in software engineering. At the end of the week, each participant sent us their chat logs and filled out an exit survey.

We qualitatively analyzed 180 dialogues that consist of 580 prompts regarding the nature and purpose of interactions, dialogue types, flow of conversation, and other attributes using interpretative phenomenological analysis~\cite{daSilva2015being} which is a research method that allowed us to capture patterns in personal experiences and better understand the characteristics of the dialogues.


In addition, the exit survey provided valuable insights into subjective assessments of productivity with ChatGPT, as well as for which purposes the chatbot was perceived as useful.

In short, this article combines our findings from this earlier study with established knowledge on human-to-human and human-to-chatbot interactions~\cite{clark2019makes}, in order to provide context for how interactions with LLMs differ from conversations between humans on the one hand, and conversations between humans and traditional NLU-based chatbots on the other.

\section{Examples of Conversations in Software Engineering}

In this article, we particularly contrast three flavors of software engineering conversations --- conversations between humans, between humans and more well-established NLU-based chatbots (henceforth called NLU-chatbots) such as GitHub bot on Slack\footnote{\url{https://slack.github.com}}, and, more recently, with LLM-powered chatbots such as ChatGPT\footnote{\url{https://chat.openai.com/}} and Bard\footnote{\url{https://bard.google.com}}. Depending on the agents involved, conversations can have different purposes and happen during planning and syncing, pair programming, requesting assistance, or just chit-chatting.

\begin{figure*}[!ht]
    \centering
    \includegraphics[width=\textwidth]{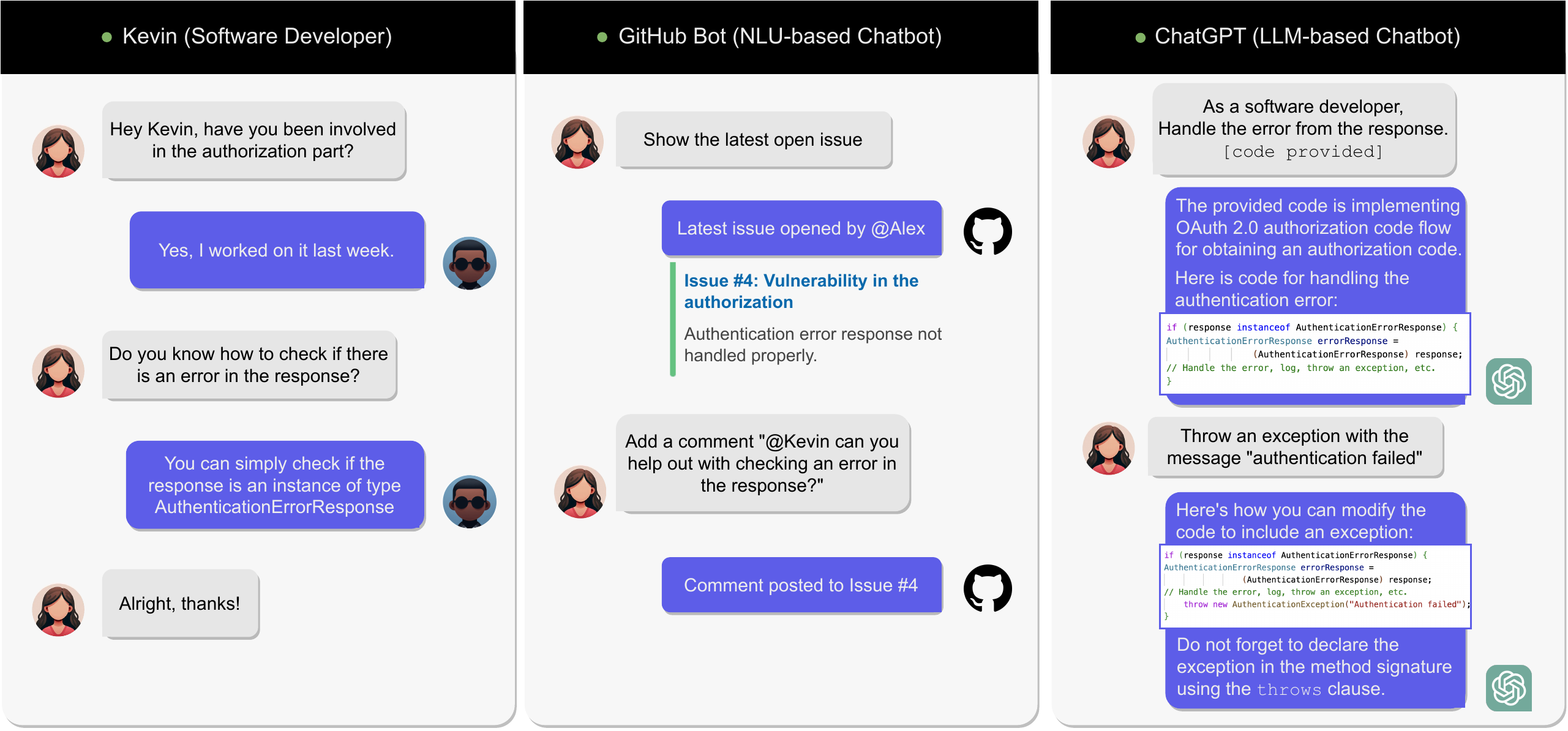}
    \caption{Three example conversations between a software developer and a fellow software developer (left), an NLU-based chatbot (middle), and an LLM-based chatbot (right).}
    \label{fig:example}
    \vspace{1cm}
    \vspace{-5px}

\end{figure*}

Let us consider an example: Alex (a junior software developer) recently joined a development team that follows an Agile process and commits daily to GitHub. The team works with authentication and uses Java, which Alex is not fully experienced in, but she knows that colleagues such as Kevin  (a senior Java developer), can help her when needed. Alex faces a problem where she needs to handle the response from an authenticator service if it returns an error. 

Figure \ref{fig:example} illustrates three possible conversations with three different types of agents.
In the first conversation, Alex seeks help from her colleague, Kevin, via the company Slack. In the conversation, Alex's first task is to establish a mutual understanding and validate that Kevin is the right colleague to ask before proceeding to ask for help. As a response, Kevin explains the basic idea Alex can follow to solve the problem,  without necessarily providing complete solutions (as a tool such as ChatGPT would). However, Kevin is unlikely to invent an entirely irrelevant solution, and Alex will generally have high trust in the correctness of Kevin's answer. Additionally, this interaction may have social benefits for the team, increasing rapport between Kevin and his new colleague.

In the second conversation, Alex uses a traditional chatbot, such as GitHub Bot, to help with her issue. Despite also supporting a conversational interface, such tools provide a much narrower range of support in comparison to either asking a colleague or using an LLM-based tool. Hence, Alex is not able to directly ask for a solution to her problem. However, she can use the chatbot to retrieve relevant information or post new entries e.g., on the project issue tracker. Alex's trust in the retrieved information will generally be high, but if the chatbot does not support her specific request she is out of luck. Further, in comparison to the other types of conversations, Alex will need to be comparatively formulaic in her request, as the chatbot's capabilities to parse natural language will be limited in comparison to the other options.

In the third conversation, Alex uses ChatGPT as it is highly accessible and readily available (unlike, for example, Kevin, who may be unavailable for a conversation). Alex starts by providing context to her question. In comparison to the first conversation, Alex needs to be much more deliberate about which context to provide and in which form. ChatGPT then explains the code provided and proposes a solution. Alex is able to ask follow-up questions and build on the previous solution, much like a conversation with another human. However, unlike conversing with a human, Alex needs to consider what context to provide and how to frame her question (often referred to as prompt engineering), and she needs to carefully check the correctness of the provided solution to identify possible hallucinations. Nevertheless, ChatGPT will (ideally) provide a working code, which Alex can use or adapt quickly.

It is evident that, despite all three scenarios involving Alex "chatting" via natural language, the actual context as well as the style of conversation (and what Alex can expect to get out of it) varies dramatically. In the following, we explore these differences more systematically.

\section{A Comparison of Conversations}

We use the framework first described by Clark et al. \cite{clark2019makes} to compare and contrast the three styles of conversations summarized in Table \ref{tab:chars}. This framework entails attributes of conversations between humans and traditional chatbots. We focus on the following attributes: (i) purpose refers to the reason the conversation happens; (ii) understanding of scope as the ability to comprehend the context of a conversation; (iii) listening - comprehending the content of a conversation; (iv) trustworthiness as the ability to provide reliable outcomes; and (v) the use of humour by lightening the conversation and making it more amusing.

\begin{table*}[!th]
    \centering
    \caption{Summary of the conversation attributes between software developers and (i) other developers, (ii) NLU-based chatbots and (iii) LLM-based chatbots }
    \label{tab:chars}
    \begin{tabularx}{\linewidth}{lXXX}
        
    \toprule
         & \textbf{Human Software Developers} & \textbf{NLU-based Chatbots} & \textbf{LLM-based Chatbots} \\
    \midrule
        \textbf{Purpose}  & Social, general guidance, training & Basic information retrieval, simple automation & General guidance, training, artifact generation and manipulation \\
        \midrule
        \textbf{Understanding of Scope} & Mutual understanding & Fixed customization & Dynamic customization \\
        \midrule
        \textbf{Listening} &  Body language (Active) and knowledge (Accurate) & Acknowledgement (Active) and intent classification (Accurate) & Query summary (Active) and knowledge (Accurate) \\
        \midrule
        \textbf{Trustworthiness} & Shared experiences and previous interactions & Performance and efficiency & Meeting expectations and transparency \\
        \midrule
        \textbf{Use of Humour} & Common & Not applicable &  On-demand\\
    \bottomrule
    \end{tabularx}

\end{table*}

\subsection{Purpose of interaction}

\begin{coloredframe}{boxcolor}
The purpose is the outcome that one expects from a conversation that helps achieve a bigger goal. Using our example in Figure \ref{fig:example}, we explain the different purposes.
\end{coloredframe}

Conversations among human developers can have social or more goal-oriented purposes. While sometimes developers may ask others to outright perform a task (e.g., delegation), many conversations are arguably targeted at receiving information or training \cite{clark2019makes}. This is illustrated in our example when Alex asks Kevin \textit{how to} check for an error to get the needed information and guidance to complete the error-checking task rather than ask him for the complete syntax. It should be noted that there is often a social benefit to the conversation, even if the conversation's original purpose is a more technical one. 

On the other hand, conversations with NLU-based chatbots are limited to delegating or performing a set of tasks (usually straightforward automation tasks) and general information retrieval. More personalized guidance and training are not possible in many development-related NLU-chatbots. In the example conversation with GitHub bot, Alex tries to achieve her goal of getting a solution by utilizing the limited queries that she can use to interact with GitHub bot, so she asks the bot to assign the open issue to Kevin.

LLM-based chatbots support a much wider range of purposes. Our data reveals that developers ask ChatGPT general questions (123, or 68\%, of interactions), but also saw them use the LLM to generate or manipulate concrete code artifacts (32\%). In the conversation between Alex and ChatGPT, Alex's main goal is to handle an error from the HTTP response in her Java code. To achieve this goal, she has three options as purposes of the conversation: She can get some \textit{training} from ChatGPT to learn about error handling and HTTP responses in Java, she can get general \textit{information} on how to solve her problem and what logic to follow (similar to what a human colleague would likely provide), or she can ask for \textit{artifact manipulation} and get the specific code solution.
Alex decides to go with the third option as it makes her reach her goal faster.

\subsection{Understanding of Scope}

\begin{coloredframe}{boxcolor}
Individuals use some information, i.e. the scope, as a foundation of a conversation that ensures that it flows towards the intended purpose. The information can take the form of assumptions or be explicitly shared in the conversation that the conversation is to be built on.
\end{coloredframe}

Conversations with human developers are built on mutual understanding. Before Alex asks Kevin for help, she makes sure he is familiar with the code on which the question is based. Since Kevin made it clear that he had seen the code, a common ground is established and the conversation carries on. Note that Alex does not need to share more information regarding the project (e.g., programming language, or project dependencies) since she assumes that Kevin knows it already given that they work in the same team.
However, instead of mutual understanding, NLU-chatbots are customized during the design phase to hold conversations that align with the user's preferences, history, and specific context. The NLU-chatbot that Alex uses (i.e., GitHub bot) is customized to the context of the project Alex works on. Therefore, when Alex asks to show the latest issue, it displays the issue for her project. Technical personalization takes the place of a socially constructed common ground.

While LLM-chatbots can be customized during design or deployment, they also provide the option for customization during the interaction, e.g., through prompt engineering. Developers provide a context or a perspective that the LLM-chatbot should consider. In our data, 62 prompts (34\%) included contextual information such as domain-specific knowledge, production code, etc. In our example, Alex steers the conversation toward getting development-related assistance when asking ChatGPT to take the role of a software developer. Alex also provides a context (that is, her code) on which she expects the rest of the conversation to be based.

\subsection{Listening}

\begin{coloredframe}{boxcolor}
Listening is the act of actively receiving and interpreting the data shared within a conversation. The data include underlying intention, information shared, and similar.
\end{coloredframe}

There are two aspects of listening: active listening and accurate listening. In active listening, the listener focuses on receiving the information shared, while accurate listening concerns interpreting the information correctly. While both aspects are needed in a conversation, they are expressed differently in different conversations.

Active listening is needed in a software organization and is applied in common activities such as daily stand-ups. When one developer verbalizes their progress to other fellow developers who show engagement through body language and maintaining eye contact, even if a solution cannot be provided, active listening helps developers organize their thoughts.
On the other hand, accurate listening requires the developer to be familiar with the discussed query and have enough knowledge and expertise to interpret the query, which is why Alex chose to ask Kevin who is a Java expert. 

For NLU-chatbots, active listening is present in acknowledgments of the developers' query (GitHub bot posts the comment and acknowledges it to Alex), whereas accurate listening is controlled by the NLU component of the chatbot. When the NLU accurately performs intent classification (predicts the user's intention) and entity extractions (extracts relevant information), it decreases the occurrences of intent misclassification and unnecessary clarification questions, hence, maintaining a good conversation flow. For example, the conversation between Alex and GitHub bot is short and effective since GitHub bot could understand Alex's intent (i.e., adding a comment) and the correct entities (e.g., issue \#4).

Accurate listening also applies to LLM-chatbots, but unlike NLU-chatbots, they do not focus on understanding the intent and entities rather than using the wide knowledge that it was trained on and the transformer architecture \cite{han2021transformer} to understand the language structures, syntax, and contextual connections in the prompt and generate a contextually-relevant response. Another feature of modern LLM-based chatbots is their application of active listening where they convey their understanding and then respond. In fact, 30 prompts (16\%) distributed among 12 participants offered feedback to ChatGPT on whether its recommendations were meeting their expectations, i.e., whether they felt like ChatGPT actively listened to them. While the percentage of prompts was low, that was done by half of our participants which, we argue, indicates a similar pattern to conversations between humans. When Alex asked ChatGPT to handle the error and provided her code, ChatGPT started explaining how it interpreted the code in terms of its functionality and then provided the code solution.

\subsection{Trustworthiness}

\begin{coloredframe}{boxcolor}
Trust is the confidence that the result will be beneficial, which enables the conversation to begin and continue.

\end{coloredframe}

Trust in human developers is built through time, by accumulating information and experiences about the developer, through interactions and shared moments. The interactions can be during team discussions, pair programming sessions, or personal consultations. Alex trusts Kevin enough to ask him since she knows that he always welcomes her questions and she knows that he is knowledgeable in the area she needs help in and can provide useful advice and guidance. Alex also trusts Kevin's answers when he says that he worked on the code, which supports the flow of the conversation and allows Alex to move to discuss her problem.

When it comes to tools, trust also needs to be established similarly \cite{PRZEGALINSKA2019785}. While the outcome of conversations with humans can be influenced by unpredictable factors (such as the time and person's mood), tool performance tends to be more consistent. 
NLU-chatbots are trusted when they can perform their tasks accurately, for instance, displaying the correct issue requested by Alex. Developers trust NLU-chatbots when using them to automate tasks and retrieve information becomes useful and efficient. The reputation of the tool's maker also plays a relevant role in establishing trust.

For LLM-chatbots, accuracy is hard to measure, especially for complex queries. Instead, developers trust chatbots that meet their expectations. In our observational study, developers who expected ChatGPT to be impeccable judging from how smart it sounds when generating a response, ended up losing trust when ChatGPT hallucinates, whereas developers with more modest expectations often found themselves positively surprised~\cite{khojah2024beyond}. In their exit surveys, 8 practitioners (33\%) stated having little to no trust in ChatGPT's answers, whereas the remainder reported trusting ChatGPT's answer (between moderate and some trust).
Another expectation is transparency, developers expect ChatGPT to communicate its level of confidence in the answer it provides (a goal that ChatGPT rarely meets, as the tool is reputed for being overconfident) and hope to see a source of the shared information (which ChatGPT, again, cannot deliver accurately). Hence, developers have good reasons to have lower trust in LLM-chatbots than in the alternatives, requiring them to carefully cross-check the outcome of interactions.

Particularly in the LLM context, a second trust angle is related to privacy. Where both humans as well as NLU-chatbots are generally considered unproblematic in a privacy sense, many privacy challenges have recently been raised towards tools such as ChatGPT. Lack of clarity on what commercial LLMs do with proprietary information and artifacts that get passed as context is a critical issue that is right now still hampering their more widespread adoption in industry.

\subsection{Humour}

\begin{coloredframe}{boxcolor}
Humour refers to the ability to add lightness or enjoyment in a dialogue making it more memorable and enhancing engagement between the parties. 
\end{coloredframe}

Humour is a quality that is mostly acceptable from humans in specific scenarios. It serves social purposes to soften the seriousness of some topics in a conversation.
Even in a technical conversation, humour can be used by Kevin to establish rapport and help Alex digest difficult information. Recent work in software engineering has shown how humour makes developers more engaged in their tasks and even helps understand complex programming tasks \cite{tiwari2024humor}.

Neither NLU-based nor LLM-based chatbots effectively make use of humour the way an emphatic human would, and if these tools employ humour it is sometimes perceived as offputting. However, LLM-based chatbots are able to reply with a certain sense of humour (but need to be explicitly asked to do so).

\section{Discussion}
Next, we summarise the differences between the different types of conversations described above while discussing the main implications and lessons learned from our study.

\subsection{Developers should adapt their expectations based on who they converse with}

For example, when conversing with human developers, they can expect in-depth conversations that can involve emotions or humor, and rely on shared understanding and collaboration. While conversations with NLU-based chatbots may prioritize concise answers and accurate task automation over providing complete and elaborate answers.
In LLM-based chatbots, developers should understand that while there are many factors that affect the outcome of the LLM-chatbots (e.g., the training data and architecture of the LLM), it is very sensitive to prompts. Prompts can either yield unintended results in case of ambiguity or enable utilizing most of LLM capabilities (e.g., complete answers to complex queries) when constructed based on recommended prompt techniques. We believe that adjusting such expectations is one step towards having more productive interactions within a software team.

\subsection{Trustworthiness is the attribute that mainly determines the flow of a conversation}

Trustworthiness impacts the usefulness of the conversations \cite{clark2019makes}.
By controlling the type and amount of information that can be shared within a conversation, it determines how the outcome of the conversation will be implemented by the developer. 
For instance, when seeking guidance from a trusted colleague, a developer can confidently implement the recommendations received, knowing they are reliable.
While conversing with LLMs that are known to generate erroneous or irrelevant information, as in the case of hallucination, can lead the developer to only use the outcome as a source of inspiration without letting it be a guide for decision-making.

\subsection{LLM-based chatbots enable software developers to have more human-like conversations, but with bot-alike efficiency}

LLM-based chatbots allow software developers to engage in conversations that have similar attributes to conversations with humans, such as the ability to express their knowledge and provide guidance. 
However, these attributes extend to allow for more conversational possibilities that are inherited from bot-based interactions in general and LLM capabilities specifically. 
For example, human-to-LLM conversations can be more flexible and available compared to human-to-human conversations. LLM-based chatbots have the capacity to comprehend a wide range of topics and adapt their responses based on the context of the conversation. This flexibility allows them to cover different purposes of conversations, including providing specific artifacts to solve problems in addition to providing recommendations and general guidance.

\subsection{Conversation styles are not mutually exclusive, but rather complementary}

Communication is an essential part of software development. Researchers have been tackling this by proposing solutions for coordination and communications in software teams \cite{locococo}. In the new era of AI-driven software development (AIware), software teams are evolving into a hybrid model involving software engineers but also AI. Consequently, new challenges have emerged in regard to the adoption of NLU-chatbots \cite{abdellatif2021comparison} and more recently LLM-chatbots e.g., the challenge of \emph{``crafting effective prompts''} discussed by \cite{hassan2024rethinking}. 
With respect to the conversation attributes, one challenge of human-to-LLM conversations is establishing trust when certain criteria are missing, such as transparency. This can be mitigated by involving other communication styles where trustworthiness criteria are available, for example, with NLU-chatbots providing a confidence estimation or with human developers with previous trustworthy reputations.
Hence, different conversation styles can be combined to mitigate the limitations of individual ones and amplify their advantages.

\section{Concluding Remarks}
LLM-based chatbots are here to stay in software development organizations. With their human-like ability to inform, generate, and create, combined with the always-on availability of an IT service, they fill a complementary niche that neither pre-existing NLU-based chatbots nor human colleagues can fill. However, we argue that LLM-based chatbots are not directly replacing human co-workers, nor should they --- the social aspect of human interactions cannot be filled even by the most advanced bot. Instead, LLM-based chatbots should be seen as a new and powerful form of generic productivity tool, which can be used effectively to decrease the ever-growing mental load~\cite{rubin:16} placed on modern developers.

\section*{Acknowledgment}
This work was partially supported by the Wallenberg AI, Autonomous Systems and Software Program (WASP) funded by the Knut and Alice Wallenberg Foundation. 

\bibliographystyle{ieeetr}
\bibliography{bib.bib}

\begin{thebibliography}{10}

\bibitem{ross2023programmer}
S.~I. Ross, F.~Martinez, S.~Houde, M.~Muller, and J.~D. Weisz, ``The programmer’s assistant: Conversational interaction with a large language model for software development,'' in {\em Proceedings of the 28th International Conference on Intelligent User Interfaces}, IUI '23, (New York, NY, USA), p.~491–514, Association for Computing Machinery, 2023.

\bibitem{ipek2023applications}
I.~Ozkaya, ``Application of large language models to software engineering tasks: Opportunities, risks, and implications,'' {\em IEEE Software}, vol.~40, no.~3, pp.~4--8, 2023.

\bibitem{clark2019makes}
L.~Clark, N.~Pantidi, O.~Cooney, P.~Doyle, D.~Garaialde, J.~Edwards, B.~Spillane, E.~Gilmartin, C.~Murad, C.~Munteanu, {\em et~al.}, ``What makes a good conversation? challenges in designing truly conversational agents,'' in {\em Proceedings of the 2019 CHI conference on human factors in computing systems}, CHI '19, (New York, NY, USA), pp.~1--12, Association for Computing Machinery, 2019.

\bibitem{erlenhov2019current}
L.~Erlenhov, F.~G. de~Oliveira~Neto, R.~Scandariato, and P.~Leitner, ``Current and future bots in software development,'' in {\em 2019 IEEE/ACM 1st International Workshop on Bots in Software Engineering (BotSE)}, pp.~7--11, IEEE, 2019.

\bibitem{wessel2018power}
M.~Wessel, B.~M. De~Souza, I.~Steinmacher, I.~S. Wiese, I.~Polato, A.~P. Chaves, and M.~A. Gerosa, ``The power of bots: Characterizing and understanding bots in oss projects,'' {\em Proceedings of the ACM on Human-Computer Interaction}, vol.~2, no.~CSCW, pp.~1--19, 2018.

\bibitem{abdellatif2020msrbot}
A.~Abdellatif, K.~Badran, and E.~Shihab, ``Msrbot: Using bots to answer questions from software repositories,'' {\em Empirical Software Engineering}, vol.~25, pp.~1834--1863, 2020.

\bibitem{lee2021don}
M.~Lee and S.~Lee, ``“i don't know exactly but i know a little”: Exploring better responses of conversational agents with insufficient information,'' in {\em Extended Abstracts of the 2021 CHI Conference on Human Factors in Computing Systems}, (New York, NY, USA), pp.~1--5, Association for Computing Machinery, 2021.

\bibitem{yang2019understanding}
X.~Yang, M.~Aurisicchio, and W.~Baxter, ``Understanding affective experiences with conversational agents,'' in {\em proceedings of the 2019 CHI conference on human factors in computing systems}, (New York, NY, USA), pp.~1--12, Association for Computing Machinery, 2019.

\bibitem{khojah2024beyond}
R.~Khojah, M.~Mohamad, P.~Leitner, and F.~G. de~Oliveira~Neto, ``Beyond code generation: An observational study of chatgpt usage in software engineering practice,'' in {\em Proceedings of the 2024 Fundamentals of Software Engineering Conference (FSE)}, (New York, NY, USA), Association for Computing Machinery, 2024.

\bibitem{khojah2023replication}
R.~Khojah, M.~Mohamad, P.~Leitner, and F.~G. de~Oliveira~Neto, ``{Package for An Observational Study of ChatGPT Usage in Software Engineering Practice},'' Sept. 2023.

\bibitem{daSilva2015being}
C.~da~Silva~Cintra and R.~A. Bittencourt, ``Being a pbl teacher in computer engineering: An interpretative phenomenological analysis,'' in {\em 2015 IEEE Frontiers in Education Conference (FIE)}, pp.~1--8, IEEE, 2015.

\bibitem{han2021transformer}
K.~Han, A.~Xiao, E.~Wu, J.~Guo, C.~Xu, and Y.~Wang, ``Transformer in transformer,'' {\em Advances in Neural Information Processing Systems}, vol.~34, pp.~15908--15919, 2021.

\bibitem{PRZEGALINSKA2019785}
A.~Przegalinska, L.~Ciechanowski, A.~Stroz, P.~Gloor, and G.~Mazurek, ``In bot we trust: A new methodology of chatbot performance measures,'' {\em Business Horizons}, vol.~62, no.~6, pp.~785--797, 2019.
\newblock Digital Transformation and Disruption.

\bibitem{tiwari2024humor}
D.~Tiwari, T.~Toady, M.~Monperrus, and B.~Baudry, ``With great humor comes great developer engagement,'' in {\em Software Engineering in Society (ICSE-SEIS)}, (Piscataway, NJ), IEEE Press, 2024.

\bibitem{locococo}
M.~Mohamad, G.~Liebel, and E.~Knauss, ``Loco coco: Automatically constructing coordination and communication networks from model-based systems engineering data,'' {\em Information and Software Technology}, vol.~92, pp.~179--193, 2017.

\bibitem{abdellatif2021comparison}
A.~Abdellatif, K.~Badran, D.~E. Costa, and E.~Shihab, ``A comparison of natural language understanding platforms for chatbots in software engineering,'' {\em IEEE Transactions on Software Engineering}, vol.~48, no.~8, pp.~3087--3102, 2021.

\bibitem{hassan2024rethinking}
A.~E. Hassan, D.~Lin, G.~K. Rajbahadur, K.~Gallaba, F.~R. Cogo, B.~Chen, H.~Zhang, K.~Thangarajah, G.~A. Oliva, J.~Lin, {\em et~al.}, ``Rethinking software engineering in the era of foundation models: A curated catalogue of challenges in the development of trustworthy fmware,'' 2024.

\bibitem{rubin:16}
J.~Rubin and M.~Rinard, ``The challenges of staying together while moving fast: An exploratory study,'' in {\em Proceedings of the 38th International Conference on Software Engineering}, ICSE '16, (New York, NY, USA), p.~982–993, Association for Computing Machinery, 2016.

\end{thebibliography}
\end{document}